\documentclass[12pt]{article}
\usepackage{amssymb,amsmath,amsfonts,amsthm}
\usepackage{graphicx}
\usepackage{hyperref}
\usepackage{cite}
\usepackage{multicol}
\usepackage{subfigure}
\usepackage{epsf}

\hoffset= -2cm \voffset= -2cm \vspace{2cm}

\newcommand{\be}{\begin{equation}}
\newcommand{\ee}{\end{equation}}
\newcommand{\bea}{\begin{eqnarray}}
\newcommand{\eea}{\end{eqnarray}}

\begin{document}
\title{Axial anomaly and mixing: from real to highly virtual photons}

\date{}
\author{Yaroslav~N.~Klopot$^a$\footnote{ On leave from  Bogolyubov Institute for Theoretical Physics, Kiev,
Ukraine}\;\footnote{{\bf e-mail}: klopot@theor.jinr.ru.} ,\,
        Armen~G.~Oganesian$^{a,b}$\footnote{{\bf e-mail}: armen@itep.ru}\;\, and \
        Oleg~V.~Teryaev$^a$\footnote{{\bf e-mail}: teryaev@theor.jinr.ru}}
\maketitle
\begin{center}
{$^{a}$\em Bogoliubov Laboratory of Theoretical Physics,\\ Joint Institute for Nuclear Research,\;\;
\\Joliot-Curie 6, Dubna 141980, Russia.\\
 $^{b}$Institute of Theoretical and Experimental Physics,\;\;\\B. Cheremushkinskaya 25, Moscow 117218,
 Russia}
\end{center}
\vspace{2cm}

\begin{abstract}

The relation for transition form factors of $\eta$ and $\eta'$ mesons is obtained by combining the exact nonperturbative QCD  sum rule, following from the dispersive representation of axial anomaly, and quark-hadron duality. It is valid at all virtual photon momenta and  allows one to express the transition
form factors entirely in terms of meson decay constants. This relation is in a good agreement with experimental data.

\end{abstract}
\newpage
\section{Introduction}

Axial anomaly \cite{Bell:1969ts,Adler:1969gk} is known to be a fundamental notion of nonperturbative QCD and
hadronic physics. Usually it is considered for the case of real photons, however, the dispersive form of
axial anomaly \cite{Dolgov:1971ri}  can be considered for virtual photons
also \cite{Horejsi:1985qu,Horejsi:1994aj,Veretin:1994dn}. It leads to an exact anomaly sum rule (ASR) which
does not have perturbative corrections due to Adler-Bardeen theorem \cite{Adler:1969er} as well as
nonperturbative QCD corrections due to 't~Hooft consistency principle.  Recently,  this sum rule was applied to the analysis of pion transition form factors \cite{Klopot:2010ke} which allowed to validate  the
interpolation formula  for pion transition form factor \cite{Brodsky:1981rp,Radyushkin:1995pj}. This form
factor attracted much attention because of unexpected and provocative data of BABAR collaboration
\cite{Aubert:2009mc}. These data were followed by a stream of theoretical papers, that in particular
questioned the  QCD factorization \cite{Radyushkin:2009zg,Polyakov:2009je}, provided the detailed analysis of
perturbative and nonperturbative QCD corrections \cite{Chernyak:2009dj,Agaev:2010aq,Mikhailov:2010ud,Bakulev:2011rp} and suggested various model
approaches  \cite{Bystritskiy:2009bk,Dorokhov:2010bz,Dorokhov:2010zz,Noguera:2010fe,Pham:2011zi}. In our
anomaly-based analysis \cite{Klopot:2010ke} it  was found that  the BABAR puzzle may indicate an existence of
small nonperturbative correction to continuum which must be  compensated by significant correction to
transition form factor.

Recently, the BABAR collaboration extended the analysis and presented the data for  $\eta$ and $\eta'$ meson
transition form factors \cite{:2011hk}. These data motivated several recent papers
\cite{Kroll:2010bf,Dorokhov:2011zf,Brodsky:2011yv,Brodsky:2011xx,Zuo:2011sk}.

In this work we analyze the $\eta$ and $\eta'$   transition form factors by means of generalized ASR to
include the effects of meson mixing. Using the dispersive representation of axial anomaly the particular
(octet) combination of meson transition form factors is expressed in terms of meson decay constants only.
This expression is in a good agreement with experimental data in the whole range from real to highly virtual
photons.

\section{Anomaly sum rule in octet channel}
In this section we briefly remind the dispersive representation for axial anomaly \cite{Dolgov:1971ri} (see
also \cite{Ioffe:2006ww} for a review) and derive anomaly sum rule for the octet channel of axial current.

The VVA triangle graph correlator
\be \label{VVA}
T_{\alpha \mu\nu}(k,q)=\int
d^4 x d^4 y e^{(ikx+iqy)} \langle 0|T\{ J_{\alpha5}(0) J_\mu (x)
J_\nu(y) \}|0\rangle \ee

\noindent contains axial current $J_{\alpha5}$ and two
vector currents $J_{\mu} = (e_u\bar{u}\gamma_\mu u +e_d\bar{d}\gamma_\mu d+e_s\bar{s}\gamma_\mu s)$;  $k,q$
are momenta of photons.  The octet component of axial current which is relevant for us explicitly reads:
$J^{(8)}_{\alpha5}=\frac{1}{\sqrt{6}}(\bar{u}\gamma_\alpha \gamma_5 u +\bar{d}\gamma_\alpha \gamma_5 d -
2\bar{s}\gamma_\alpha \gamma_5 s)$.

It is convenient to write the tensor decomposition  \cite{Rosenberg:1962pp,Eletsky:1982py,Radyushkin:1996tb}
of correlator (\ref{VVA}) in a form:

\begin{eqnarray}
\label{eq1} \nonumber T_{\alpha \mu \nu} (k,q) & = & F_{1} \;
\varepsilon_{\alpha \mu \nu \rho} k^{\rho} + F_{2} \;
\varepsilon_{\alpha \mu \nu \rho} q^{\rho}
\\ \nonumber
& & + \; \; F_{3} \; k_{\nu} \varepsilon_{\alpha \mu \rho \sigma}
k^{\rho} q^{\sigma} + F_{4} \; q_{\nu} \varepsilon_{\alpha \mu
\rho \sigma} k^{\rho}
q^{\sigma}\\
 & & + \; \; F_{5} \; k_{\mu} \varepsilon_{\alpha \nu
\rho \sigma} k^{\rho} q^{\sigma} + F_{6} \; q_{\mu}
\varepsilon_{\alpha \nu \rho \sigma} k^{\rho} q^{\sigma},
\end{eqnarray}
where the coefficients $F_{j} = F_{j}(k^{2}, q^{2}, p^{2}; m^{2})$, $p
= k+q$, $j = 1, \dots ,6$ are the corresponding Lorentz invariant
amplitudes constrained by current conservation and Bose symmetry.  Note, that the latter includes the interchange $\mu \leftrightarrow \nu, k \leftrightarrow q$ in the tensor structures and $k^2 \leftrightarrow q^2$ in the arguments of scalar functions $F_{j}$.

In this paper we are interested in the case of one real ($k^2=0$) and one virtual photon ($Q^2=-q^2>0$). Then for the invariant amplitude $F_3 - F_6$ the anomaly sum rule (ASR) takes the form \cite{Horejsi:1994aj}:
\begin{equation}
\label{ASR} \int_{4m^{2}}^{\infty} A_{3a}(t;q^{2},m^{2}) dt =
\frac{1}{2\pi}N_c C^{(a)} \;,
\end{equation}
where $N_c=3$ is a number of colors and \begin{align} A_{3a}& = \frac{1}{2}Im (F_3-F_6),  \\
C^{(3)}&=\frac{1}{\sqrt{2}}(e_u^2-e_d^2)=\frac{1}{3\sqrt{2}},  \\
C^{(8)}&=\frac{1}{\sqrt{6}}(e_u^2+e_d^2-2e_s^2)=\frac{1}{3\sqrt{6}}. \end{align}

The ASR (\ref{ASR}) is an exact relation, i.e. it does not have  perturbative corrections  to the integral
\cite{Adler:1969er} as well as it does not have any nonperturbative
corrections too (as it is expected from 't Hooft's principle).
Another important property of this relation is that it holds for an arbitrary quark mass $m$ and for any
$q^{2}$.

Note, that one can also write out the similar relation with
\begin{align}
C^{(0)}&=\frac{1}{\sqrt{3}}(e_u^2+e_d^2+e_s^2)=\frac{2}{3\sqrt{3}},
\end{align}

\noindent for the singlet component of the axial current. However, in this case the absence of corrections is not guaranteed and one should explicitly take into account the gluonic anomaly. So we will not consider the
singlet channel in this paper concentrating on the octet one.

Saturating the  l.h.s. of the 3-point correlation function (\ref{VVA}) with the resonances and singling out
their contributions to  ASR (\ref{ASR}) we get the sum of resonances with appropriate quantum numbers:

\begin{eqnarray} \label{GQHD}
f^8_\eta F_\eta + f^8_{\eta'} F_{\eta'}+ (other \;resonances) = \nonumber \\ \int_{4m^{2}}^{\infty} A_{3a}(t;q^{2},m^{2})
dt=\frac{1}{2\pi}N_c C^{(8)}.
\end{eqnarray}

\noindent Here the projections of the axial current $J^{(a)}_{5\alpha}$ onto
one-meson  states $M (= \eta, \eta')$  define the coupling (decay) constants $f^a_M$:
\be \langle 0|J^{(a)}_{\alpha 5}(0) |M(p)\rangle=
i p_\alpha f^a_M \;, \ee
while the form factors $F_{M\gamma}$ of the transition $\gamma\gamma^* \to M$  are defined by the matrix elements:

\be \int d^{4}x e^{ikx} \langle M(p)|T\{J_\mu (x) J_\nu(0)
\}|0\rangle = \epsilon_{\mu\nu\rho\sigma}k^\rho q^\sigma
F_{M\gamma} \;. \ee

The relation (\ref{GQHD}) is exact and expresses the global  duality between hadrons and quarks.
Nevertheless, to analyze the hadron properties one should additionally implement the local quark-hadron
duality hypothesis. Taking into account large $\eta-\eta'$ mixing, one can express the spectral
function $A_{3a}(s,Q^2)$ in form of ``two resonances+continuum'':

\begin{eqnarray} \label{QHD}
A_{3a}\left(s,Q^2\right)= \pi f^8_{\eta}\delta(s-m_\eta^2)
F_{\eta \gamma}\left(Q^2\right) + \nonumber\\ \pi f^8_{\eta'}\delta(s-m_{\eta'}^2)
F_{{\eta'} \gamma}\left(Q^2\right) + A^{QCD}_{3a}\theta(s-s_0).
\end{eqnarray}
Here $s_0$ is a continuum threshold and $A^{QCD}_{3a}$ at one-loop level is
\be A^{QCD}_{3a}=\frac{1}{2\pi\sqrt{6}}\frac{Q^2}{(s+Q^2)^2}. \label{A_QCD}
\ee
\noindent As it was shown in \cite{Jegerlehner:2005fs} there is no two-loop $\alpha_s$ corrections to this
expression.

Analyzing (\ref{QHD}), (\ref{A_QCD}) one should note that the particles with nonzero two-photon decays cannot be included in the continuum as it vanishes at $Q^2=0$, so they should be taken into account explicitly in the ASR.  For heavy mesons  the corresponding coupling constants should  be suppressed
\cite{Ioffe:2007eg,Klopot:2009cm} at least as $(m_\eta/m_{res})^2$  which follows from the conservation of
axial current $J_{\mu5}^{(8)}$ in the chiral limit (if only strong interaction is taken into account). That is why  we restrict ourselves only to $\eta$ and $\eta'$ mesons.
The ASR for the  octet  channel then reads:

\begin{equation}\label{ASR8}
\pi f_\eta^8 F_{\eta \gamma}(Q^2)+\pi f_{\eta'}^8 F_{{\eta'}
\gamma}(Q^2)=\frac{1}{2\pi\sqrt{6}}\frac{s_0}{Q^2+s_0}.
\end{equation}

Let us stress that this relation is correct for all $Q^2$ due to the absence of the corrections to the
$Im(F_3-F_6)$ \cite{Jegerlehner:2005fs} which allows to utilize the above expression for different $Q^2$.

For real photons ($Q^2=0$) the above expression coincides with the expression in  \cite{Klopot:2009cm}, which
was obtained from dispersive approach to axial anomaly in somewhat different way:

\begin{equation}\label{ASR8-0}
\pi f_\eta^8 F_{\eta \gamma}(0)+\pi f_{\eta'}^8 F_{{\eta'} \gamma}(0)=\frac{1}{2\pi\sqrt{6}},
\end{equation}
where
\be
F_{M\gamma}(0) =\sqrt{\frac{4\Gamma_{M\to 2\gamma}}{\pi \alpha^2 m^3_{M} }}.
\ee
\noindent Here $\Gamma_{M\to 2\gamma}$ and $m_M$ are two-photon decay widths and masses of $\eta,\eta'$
mesons correspondingly.

The equation (\ref{ASR8}) allows us to fix the continuum threshold $s_0$ by considering the limit $Q^2 \to
\infty$ where the QCD factorization \cite{Efremov:1979qk,Lepage:1979zb} is applicable (contrary to
ASR the exploration of this limit  in generic QCD sum rules  is obscured  by possible corrections).
The form factors at large $Q^2$ \cite{Anisovich:1996hh,Feldmann:1997vc,*Feldmann:1999uf} are:

\be
Q^2 F_{\eta\gamma}^{as}= 2(C^{(8)}f_\eta^8+ C^{(0)}f_\eta^0)\int_0^1 \frac{\phi^{as}(x)}{x} dx
\ee
\be
Q^2 F_{\eta'\gamma}^{as}= 2(C^{(8)}f_{\eta'}^8+ C^{(0)}f_{\eta'}^0)\int_0^1 \frac{\phi^{as} (x)}{x} dx
\ee

We take into account that in the limit $Q^2 \to \infty$ the light cone distribution amplitudes of both $\eta, \eta'$ mesons are described by their asymptotical form \cite{Efremov:1979qk,Lepage:1979zb}:
$\phi^{as}(x)=6x(1-x)$.

Then the ASR for the octet channel at large $Q^2$ leads to:

\be \label{s8}
s_0 = 4\pi^2((f_\eta^8)^2+(f_{\eta'}^8)^2+ 2\sqrt{2} [ f_\eta^8 f_{\eta}^0+ f_{\eta'}^8 f_{\eta'}^0]).
\ee

Substituting (\ref{s8}) into (\ref{ASR8}) we express ASR in terms of meson decay constants $f^a_M$ only,
which is our main result:

\begin{eqnarray}\label{ASR8f}
 f_\eta^8 F_{\eta \gamma}(Q^2)+ f_{\eta'}^8 F_{{\eta'} \gamma}(Q^2)= \nonumber \\
 \frac{\sqrt{\frac{2}{3}}}{4\pi^2+Q^2/((f_\eta^8)^2+(f_{\eta'}^8)^2+ 2\sqrt{2} [f_\eta^8 f_{\eta}^0+
 f_{\eta'}^8 f_{\eta'}^0])}.
\end{eqnarray}

It is instructive to compare this formula with interpolation formulas for transition form factors of $\eta$, $\eta'$ mesons proposed in \cite{Feldmann:1998yc}. The use of the proposed interpolation formulas leads to relation which is different from (\ref{ASR8f}) but coincides with it in the limit of large and small $Q^2$. Numerically, the difference is small at all $Q^2$ (the maximal difference is  $10\%$ at $\sim1$ $GeV$) which proves that the interpolation formulas are good approximations.

Let us now pass to  applications of (\ref{ASR8f}).

\section{Mixing schemes and transition form factors}

Let us analyze the ASR (\ref{ASR8f}) for different mixing schemes. It is convenient to eliminate the
dependence of the l.h.s. of (\ref{ASR8f}) on the scale of the coupling constants  by dividing both sides of
Eq. (\ref{ASR8f}) by $f_8=\sqrt{ (f_{\eta}^8)^2+(f_{\eta'}^8)^2}$. Consider the form factors  multiplied by
$Q^2$

\begin{eqnarray} \label{ASR8-plot}
\frac{Q^2}{f_8}(f_\eta^8 F_{\eta \gamma}(Q^2)+ f_{\eta'}^8 F_{{\eta'}
\gamma}(Q^2))= \nonumber \\ \frac{\frac{Q^2}{f_8}\sqrt{\frac{2}{3}}}{4\pi^2+Q^2/((f_\eta^8)^2+(f_{\eta'}^8)^2+ 2\sqrt{2} [f_\eta^8 f_{\eta}^0+ f_{\eta'}^8 f_{\eta'}^0])}
\end{eqnarray}
and introduce the matrix of decay constants in the usual way:

\be \mathbf{F}\equiv \left(\begin{array}{cc}  f_\eta^8 & f_{\eta'}^8 \\ f_\eta^0 & f_{\eta'}^0
\end{array}\right).
\ee

$\bullet \;$ We start from  the one-angle mixing scheme, with

\be \label{1ang} \mathbf{F}= \left(\begin{array}{cc} f_8 \cos \theta & f_8\sin \theta \\
-f_0\sin \theta & f_0\cos \theta  \end{array}\right).
\ee

This mixing scheme was analyzed in many papers (see e.g. \cite{Ball:1995zv} and references  therein), giving
the values of mixing angle in the range $\theta=-12^o \div -22^o$.

In this case the ASR acquires a simple form:
\be \label{ASR8-1-plot}
Q^2( F_{\eta \gamma}(Q^2)\cos \theta+  F_{\eta'\gamma}(Q^2)\sin \theta
)=\sqrt{\frac{2}{3}}\frac{Q^2}{4\pi^2f_8+Q^2/f_8},
\ee

where constant $f_8$ is defined by the anomaly sum rule at $Q^2=0$ (\ref{ASR8-0}):
\be \label{f8-1}
f_8=\frac{\alpha}{4\sqrt{6}\pi^{3/2}} \left( \sqrt{\frac{\Gamma_{\eta\to 2\gamma}}{m^3_{\eta}}}\cos\theta +
\sqrt{\frac{\Gamma_{{\eta'}\to 2\gamma}}{m^3_{\eta'}}}\sin\theta \right)^{-1}.
\ee

Thus,  (\ref{ASR8-1-plot}) and (\ref{f8-1}) determine the mixing angle in terms of physical quantities (decay widths and transition form factors).

The corresponding relation is plotted for different mixing angles in Fig.1,2.
The dots with error bars correspond to the l.h.s. of Eq. (\ref{ASR8-1-plot}), where the form factors of
$\eta, \eta' $ mesons are taken from experimental data of CLEO \cite{Gronberg:1997fj} and BABAR
\cite{:2011hk} collaborations. The  r.h.s. of  Eq. (\ref{ASR8-plot}) corresponds to the curve with the shaded stripe defined by the experimental uncertainties of meson decay widths. The slope of the straight line in the origins indicates  the value of the anomaly sum rule at $Q^2=0$. This sum rule was considered earlier \cite{Klopot:2009cm} and is in a good agreement with the experimental values of two-photon decay widths of $\eta$ and $\eta'$ mesons.

We observe that for $\theta=-14^o \div -16^o$ there is a reasonable agreement with the experimental data
\cite{Gronberg:1997fj,:2011hk}.
The corresponding value of coupling constant (\ref{f8-1}) is  $f_8=0.87  \div  0.94 f_\pi$, $f_\pi=0.13$ $ GeV$.

The slight increase at large $Q^2$ is related, in particular, by the tendency of $\eta'$ contribution to decrease at large $Q^2$. Because of the negative mixing angle this leads to the behavior in the octet channel, qualitatively resembling the pion case, which produced the BABAR puzzle.

\begin{figure}[t]
\begin{multicols}{2}
\hfill
\includegraphics[width=0.47\textwidth]{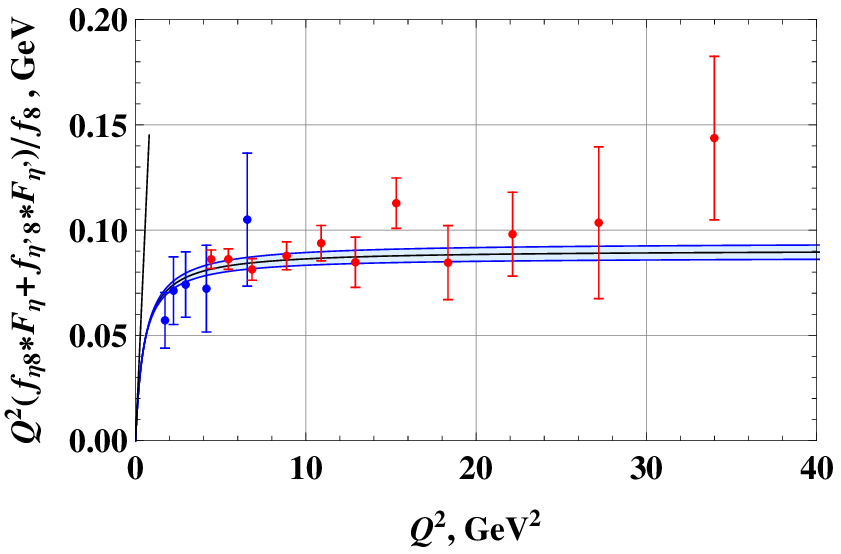}
\hfill \caption{ASR for one-angle mixing scheme (\ref{ASR8-1-plot}): $\theta=-14^o$}
\label{1ang-plot2} \hfill
\includegraphics[width=0.47\textwidth]{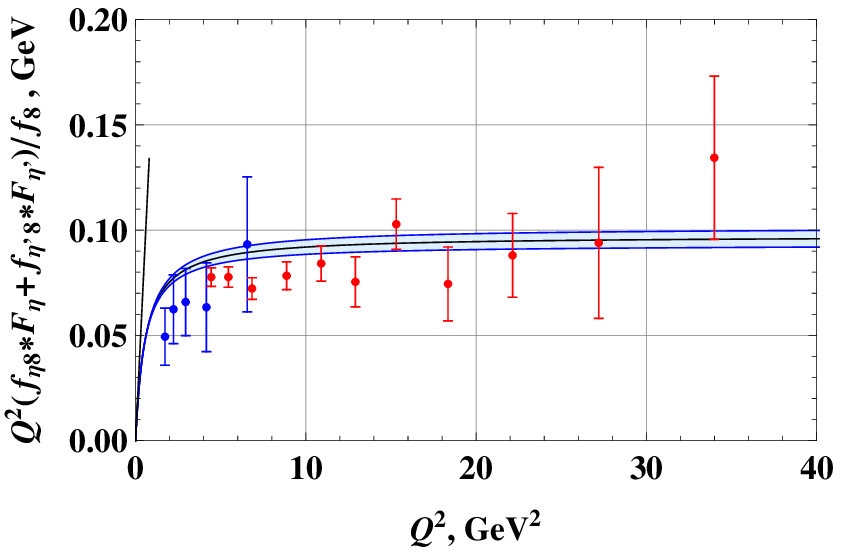}
\hfill \caption{ASR for one-angle mixing scheme (\ref{ASR8-1-plot}): $\theta=-16^o$}
\label{1ang-plot3}
\end{multicols}
\end{figure}

$\bullet \;$ Let us now discuss the mixing schemes suggested and developed in
\cite{Leutwyler:1997yr,Feldmann:1998vh,Escribano:2005qq}. These schemes parametrize the decay constants $f_M^a$ in terms of  two mixing angles $\theta_8, \theta_0$:

\be \label{2ang}\mathbf{ F}= \left(\begin{array}{cc}  f_8\cos\theta_8 & f_8\sin\theta_8 \\ -f_0\sin\theta_0 &
f_0\cos\theta_0 \end{array}\right).
\ee

Note, that this kind of matrices may appear when one considers the quark basis (see, e.g.
\cite{Feldmann:1998vh}).
For the parameters, suggested in  \cite{Feldmann:1998vh} ($f_0=1.17 f_\pi, f_8=1.26 f_\pi, \theta_0=-9.2^o,
\theta_8=-21.2^o$) the plot describing the ASR (\ref{ASR8-plot}) is shown in Fig.3.

\begin{figure}[t]
\begin{multicols}{2}
\hfill
\includegraphics[width=0.47\textwidth]{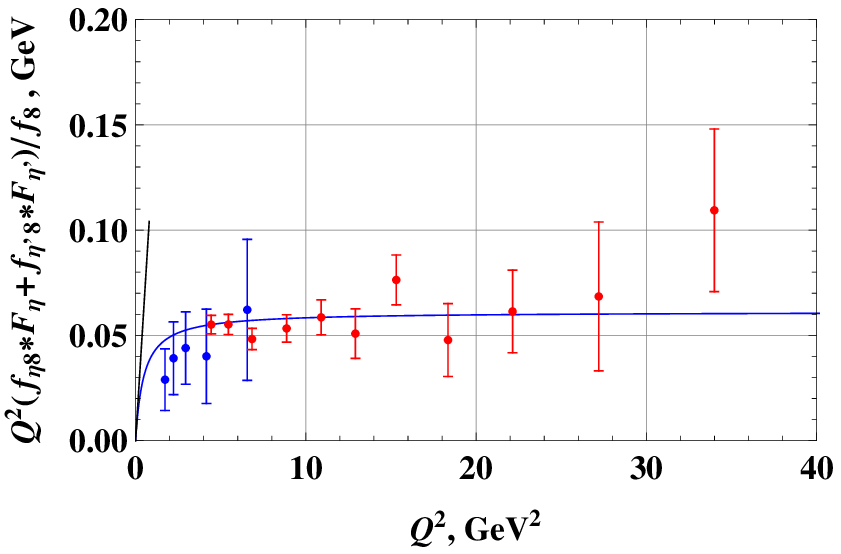}
\hfill \caption{ASR for scheme\cite{Feldmann:1998vh}: $f_0=1.17 f_\pi, f_8=1.26 f_\pi, \theta_0=-9.2^o,
\theta_8=-21.2^o$ }
\label{1ang-plot2} \hfill
\includegraphics[width=0.47\textwidth]{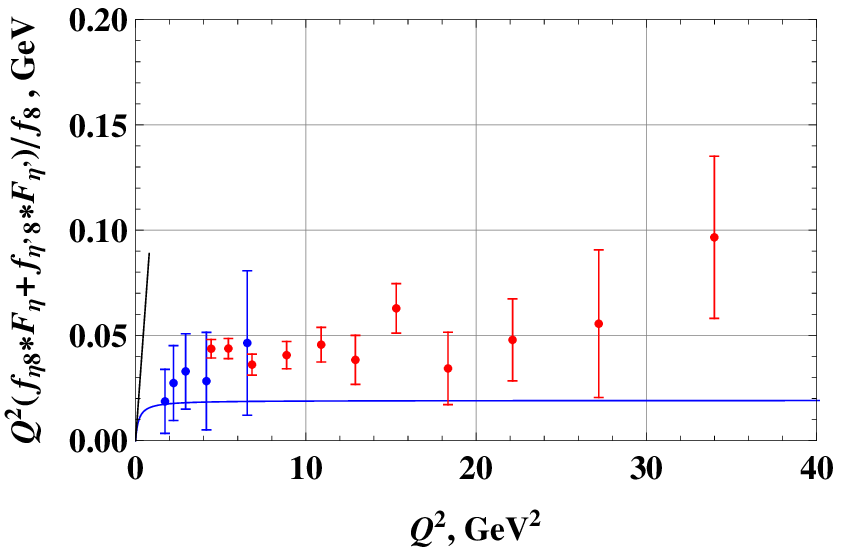}
\hfill \caption{ASR for scheme\cite{Escribano:2005qq}: $f_0=1.29 f_\pi, f_8=1.51 f_\pi, \theta_0=-2.4^o,
\theta_8=-23.8^o$ }
\label{1ang-plot3}
\end{multicols}
\end{figure}

For parameters, suggested in \cite{Escribano:2005qq} ($f_0=1.29 f_\pi, f_8=1.51 f_\pi, \theta_0=-2.4^o,
\theta_8=-23.8^o$)  the plot for  ASR  (\ref{ASR8-plot}) is shown in  Fig.4. One can see that for this
particular mixing model the agreement of ASR with the experimental data is substantially worse.

The developed approach can be used as an additional constraint on mixing parameters. It is useful to analyze the mixing schemes which take into account more than two mixing
states; the detailed analysis of these schemes is quite lengthy and will be presented elsewhere.

\section{Discussion and Conclusions}

We generalize the rigorous nonperturbative QCD approach which relies on quark-hadron duality hypothesis but  does not contain any free adjustable
parameters to the case of $\eta$ and $\eta'$ mesons where mixing is crucially important.

Combining the exact dispersive form of anomaly relation, quark-hadron duality hypothesis and asymptotic
matching with QCD factorization  we express the combination of $\eta$, $\eta'$ meson transition form factors
in terms of meson decay constants only (\ref{ASR8f}). The obtained anomaly sum rule is valid in the whole
kinematical region starting from $Q^2=0$. This ASR is quite robust and can be used as a test for different
sets of mixing parameters. Our analysis shows that for a large number of mixing schemes the ASR is in a good
agreement with experimental data (probably with the exception of parameters offered in
 \cite{Escribano:2005qq}).

At  the same time, let us note that if one estimate the value of continuum threshold $s_0$ from Eq.
(\ref{s8}) it appears to be quite small for all mixing schemes. This may reflect the contradiction of local
quark-hadron duality from one side, and anomaly from the other side. The possible resolution of such
contradiction can be due to  $1/Q^2$ suppressed nonperturbative corrections  to continuum similar to
suggested earlier \cite{Klopot:2010ke} for explanation of BABAR pion puzzle. It is significant that such
corrections are  essential for the value of $s_0$ while ASR (\ref{ASR8f}) is practically insensitive to them.
This problem as well as the consideration the singlet channel of axial current and other mixing schemes
requires a special investigation which is now in progress.

We thank  B.~L.~Ioffe,  P.~Kroll, S.~V.~Mikhailov, T.~N.~Pham, A.~V.~Radyushkin for useful discussions and elucidating comments. This  work was supported in part by RFBR (Grants 09-02-00732, 09-02-01149, 11-02-01538,11-02-01454) and by fund from CRDF  Project RUP2-2961-MO-09.

\end{document}